# Convert any android device into a programmable IoT device with the help of IoT Everywhere Framework

[1] Vishnu Joshi
*School of Cs and IT*
*Jain (Deemed-to-be) University*
Bengaluru, India
joshi.vishnu1994@gmail.com

*Abstract*—The world around us is transforming as the field of the Internet of Things is taking over the world faster than we thought. Everyone in the tech industry is building wonderful things with the help of IoT. Smartwatches, smart coffee machines, smart television, smart homes are some of the examples. Building IoT sensor modules with sensors that connect to the internet can be very intimidating for people who have just stepped into the field. Quality components and microcontrollers can be costly too. Components such as proximity sensor, humidity sensor, air pressure sensor, accelerometer, gyroscope, flashlight, microphone, speaker, gsm module, wifi module, Bluetooth modules, and many more. But to program these we need to know java or kotlin and mobile application development. With the use of the IoT Everywhere framework and Origin programming language, one can convert any Android smartphone into an IoT device. This helps students of electrical engineering to grasp the idea of programming since it provides a lot of abstraction through simple function calls it can help to introduce programming to school students, it helps students who are fascinated by IoT and who wants to learn the basic of interfacing components or sensors and helps the student who has no access to an actual personal computer learn to program.

## I. INTRODUCTION

The world around us is transforming as the field of the Internet of Things is taking over the world faster than we thought. A simple IoT prototype usually includes sensors/devices which are connected to the internet, wired/wireless access point, data processing, and data storage in the backend, and a frontend application to show all the data or analyzed data. Building nodes with sensors that connect to the internet can be very intimidating for people who have just stepped into the field. A person has to know how to program a specific microcontroller or microprocessor and interface it with sensors. He/she needs to have a rudimentary understanding of electronics. Quality development boards for microcontrollers and microprocessors and quality sensors are costly. We all do have IoT devices right in our pockets without even knowing about it. Today we cannot imagine our lives without our smartphones with us. Our phones include many components that are used without our knowledge. With the help of the IoT Everywhere framework, one can easily use the Origin programming language and interface these components easily. The IoT Everywhere framework reduces the complexity and lets the user program their mobile phones with the help of origin programming language. The Origin programming language is a simple procedural oriented programming language that is specifically made for the IoT Everywhere framework. The Origin programming language has a sugary syntax that not only helps people who are not familiar with programming but also helps children to easily learn programming. This also helps the kids in rural areas where an actual computer is not accessible to learn computer programming.

## II. ORIGIN PROGRAMMING LANGUAGE

The Origin programming language is a simple procedure-oriented programming language. It has lots of builtin functions that can be used to interface the above-mentioned components. [I]

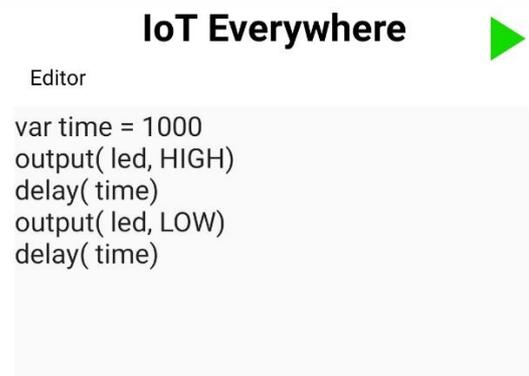

Fig 1. program to blink the flashlight once with a second delay (Screenshot from IoT Everywhere mobile app).

As you can see in Fig 1 we can see how easy it is to blink the flashlight which is equivalent to the "Hello World !!!" program in the IoT field. To achieve this a lexer, parser, and evaluator is built from scratch [1][2]. It is a simple interpreter [1][2] that is built using java to work specifically on android devices.

Origin programming language for IoT everywhere framework makes accessing smart phone's components very easy. It makes use of a combination of tailor-made builtin function calls and keywords. Every component in the device has a specific keyword assigned to it, for example, to access the output value from the accelerometer in the smartphone it has builtin function for it as shown in Fig 2. In Fig 2 accelerometerX, accelerometerY, and accelerometerZ are keywords. As seen in the above code snippet in Fig 2 the X, Y, and Z values from the accelerometer are stored in variables named x, y, and z respectively. When we look at the code snippet in Fig 1 led is a keyword. There the output function takes two parameters. The first one is the keyword which specifies which output component should be activated (speaker and led). The second parameter will be the state of that component, either HIGH or LOW, either 1 or 0 respectively. Here HIGH and LOW both are keywords.

```
var x = input( accelerometerX)
var y = input( accelerometerY)
var z = input( accelerometerZ)
```

Fig 2. Program to access values from accelerometer present in the smartphone (Screenshot from IoT Everywhere mobile app).

Making phone calls and sending messages is also very easy as shown in Fig 3 and Fig 4.

```
var num = "+911234557890"
call( num)
```

Fig 3. Program to make a phone call (Screenshot from IoT Everywhere mobile app).

```
var num = "+911234557890"
var msg = "Hello world !!"
message( num, msg)
```

Fig 4. Program to send a message (Screenshot from IoT Everywhere mobile app).

*A. Conditional Statements*

In the origin programming language, there are three types of conditional branching statements, if statement, if-else statement, and if-else-if statement. The conditional statements are simple and easy and similar to the c programming language or java programming language. The use of conditional statements is shown in Fig 5.

```
var a = input( accelerometerX)

if( a > 0){

    output( led, HIGH)

}else{

    output( led, LOW)

}
```

Fig 5. Program showing use of conditional statement (Screenshot from IoT Everywhere mobile app).

*B. Looping statements*

In Origin programming language looping statements are made simple by making use of only one statement. The use of the looping statement is shown in Fig 6, Fig 7, Fig 8 and Fig 9.

```
loop(){

    output( led, HIGH)
    wait(1000)
    output( led, LOW)
    wait(1000)

}
```

Fig 6. Program showing infinite loop (Screenshot from IoT Everywhere mobile app).

```
var i = 0

loop( i < 10 ){

    output( led, HIGH)
    wait(1000)
    output( led, LOW)
    wait(1000)
    i = i + 1

}
```

Fig 7. Program implementing while loop (Screenshot from IoT Everywhere mobile app).

```
IoT Everywhere
Editor
var i = 10

loop( i ){

   output( led, HIGH)
   wait(1000)
   output( led, LOW)
   wait(1000)

}
```

Fig 8. Program implementing for loop (Screenshot from IoT Everywhere mobile app).

```
IoT Everywhere
Editor
var arr = [ 500, 1000, 1500, 2000]
var i

loop( i in arr ){

   output( led, HIGH)
   wait(i)
   output( led, LOW)
   wait(i)

}
```

Fig 9. Program implementing for-each loop (Screenshot from IoT Everywhere mobile app).

As we can see in the code snippets in Fig 6, Fig 7, Fig 8, Fig 9, the loop statement used is the same, but based on the parameter given the interpreter decides whether it's a for loop, while loop, or for-each loop.

Since it's an IoT framework making network calls and making use of Wi-Fi and Bluetooth is an integral part of communication. Making network calls are made easy with the implementation of builtin functions such as to get, post, put, delete. Code snippet in Fig 10 shows implementation of network calls.

```
IoT Everywhere
Editor
var ssid = "hello"
var password = "world"
var conStatus = wifiConnect( ssid, password)

if( conStatus == 1 ){

   var a = json( "name", "vishnu", "age", 20 )
   var req = request("http://sampleurl.com")

   var x = input( gyroscopeX )
   addJsonElement( a, "gyroscopeX", x )

   addJson( req, a)
   var result = post( req )

   if( result == 1 ){

      output( " post request successful " )

   }else{

      output( " post request failed " )

   }

}else{

   output( "Wifi not connected" )

}
```

Fig 10. Program to implement network calls (Screenshot from IoT Everywhere mobile app).

III. CONCLUSION AND FUTURE WORK

With the use of the IoT Everywhere framework and Origin programming language, one can convert any Android smartphone into an IoT device. This helps all the people who have an interest in learning programming, especially students of electrical engineering to grasp the idea of programming, since it provides a lot of abstraction through simple function calls it can help to introduce programming to school students, students who are fascinated by IoT to learn the basic of interfacing components or sensors and to the student who has no access to an actual personal computer.

Future work on this research will include implementation of cloud connectivity, where one can write a program in Origin programming language to access cloud databases such as Firebase, Google Cloud, Amazon Web Services.